\documentclass[final,1p,times,twocolumn,authoryear]{elsarticle}

\usepackage{amssymb}
\usepackage{color}

\usepackage{amsmath}
\usepackage{hyperref}
\journal{Astroparticle Physics}

\newcommand{\mpo}{\textcolor{black}}
\newcommand{\anatoli}{\textcolor{black}}

\begin{document}

\begin{frontmatter}



\title{Production of secondary particles in heavy nuclei interactions in supernova remnants}


\author[1,2]{Maulik Bhatt}
\author[2,3,4]{Iurii Sushch}
\author[2,5]{Martin Pohl}
\ead{marpohl@uni-potsdam.de}
\author[6]{Anatoli Fedynitch}
\author[2,5]{Samata Das}
\author[2,5]{Robert Brose}
\author[7]{Pavlo Plotko}
\author[5]{Dominique M.-A. Meyer}

\address[1]{Sardar Vallabhbhai National Institute of Technology, India}
\address[2]{DESY, D-15738 Zeuthen, Germany}
\address[3]{Centre for Space Research, North-West University, 2520 Potchefstroom, South Africa}
\address[4]{Astronomical Observatory of Ivan Franko National University of L'viv, vul. Kyryla i Methodia, 8, L'viv 79005, Ukraine}
\address[5]{University of Potsdam, Institute of Physics and Astronomy, 14476 Potsdam, Germany}
\address[6]{Institute for Cosmic Ray Research, the University of Tokyo,
5-1-5 Kashiwa-no-ha, Kashiwa, Chiba 277-8582, Japan}
\address[7]{Taras Shevchenko National University of Kyiv, Ukraine}

\begin{abstract}
\mpo{Depending on their type, supernova remnants may have ejecta material with high abundance of heavy elements such as carbon or oxygen. In addition, core-collapse supernovae explode in the wind material of their progenitor star that may also have a high abundance of heavy elements.} Hadronic collisions in these enriched media spawn the production of gamma rays, neutrinos, and secondary electrons with spectra that cannot be scaled from those calculated for pp collisions, potentially leading to erroneous results. We used Monte-Carlo event generators to calculate the differential production rate of particles such as gamma rays, neutrinos, and secondary electrons for H, He, C, and O nuclei as projectiles and as target material. The cross sections and the multiplicity spectra are separately computed for each of the 16 combinations of projectile and target. We describe characteristic effects of heavy nuclei in the shape and normalization of the spectra of the various particles produced.

\end{abstract}

\begin{keyword}
Cosmic rays \sep Gamma rays \sep Abundances \sep Supernova remnants

95.85.Pw \sep 95.85.Ry \sep 98.35.Bd \sep 98.38.Mz

\end{keyword}

\end{frontmatter}


\section{Introduction}
\label{intro}
The composition of cosmic rays reflects the interactions they undergo during their propagation through the interstellar medium. It is also shaped by the particles abundance at their acceleration sites in the Galaxy. The flux of secondary cosmic rays, i.e. those that are primarily produced by spallation reactions during the propagation of highly energetic particles, can be used to determine their propagation history in our Galaxy and hence infer their source composition \citep{2016ApJ...824...16J}. It was noted more than a decade ago \citep{2007SSRv..130..439B}, that the isotopic ratios in Galactic cosmic rays at a few hundred MeV/nuc are consistent with a source composition consisting of 20\% Wolf-Rayet material \citep{1987ApJ...315..209P,1993A&A...278..406M} mixed with 80\% solar-system material. The Wolf-Rayet material is predominantly found in large superbubbles \citep{1998ApJ...509L..33H}, and it is therefore tempting to attribute a fair fraction of the Galactic cosmic-ray  production to superbubbles \citep{2005ApJ...628..738H}, where the large number of shocks may collectively accelerate plasma particles to very high energy \citep{2001AstL...27..625B,2004A&A...424..747P}.

Massive stars ($\ge8\, \rm M_{\odot}$) are characterized by their evolving surface and wind properties. These stellar objects can ultimately evolve through the so-called Wolf-Rayet phase characterized by a high mass-loss rate ($\sim 10^{-5}\, \rm M_{\odot}\, \rm yr^{-1}$) and a high terminal wind velocity~\citep{langer_araa_50_2012}. This happens after a sequence of various consecutive evolutionary stages, whose complexity is a function of the zero-age-main-sequence mass -- that of a forming high-mass stellar object when the core has reached the molecular-hydrogen dissociation temperature and the surface begins to develop a supersonic wind --, its rotational properties and/or its chemical composition \citep[see e.g.][]{2011A&A...530A.115B,2012A&A...537A.146E,2015A&A...581A..15S}. The composition of the corresponding winds reflects the distribution of heavy material produced by nuclear fusion in the stellar core and transported by convection to the upper radiative layer of the star \citep{2015A&A...581A..15S}. The interaction of stellar winds of massive stars with the ambient interstellar medium leads to the formation of wind nebulae, whose morphology depends on the properties of the stellar and ambient medium, forming, e.g., spherical wind bubbles around stars at rest~\citep{1977ApJ...218..377W} and bow shocks around runaway stars~\citep{wilkin_459_apj_1996}. 

Therefore, such enriched material is naturally found in the circumstellar nebulae generated during the evolved phase of massive stars~\citep{mackey_apjlett_751_2012,2014Natur_512_282M}. An example is the bow-shock nebula observed around the runaway red supergiant IRC-10414 discovered by N[II] line-emission excess~\citep{Gvaramadze_2013, meyer_2014a}. It is also found in the ejecta of core-collapse supernovae \citep{2013ApJ...764...21C,Vink2017}, although the uncertainties in the yields are large. A particular puzzle of cosmic-ray acceleration in enriched media is the recent detection of $^{60}$Fe at 500 MeV/nuc \citep{2016Sci...352..677B}. This unstable isotope is preferentially
produced in supernovae and not abundant in winds. Its detection indicates that within less than a few million years after production in the supernova the particles were accelerated and transported
to Earth.

Particle acceleration likely arises by diffusive shock acceleration \citep{1987PhR...154....1B}. The forward shock of core-collapse supernova remnants will first propagate through the wind of the last evolution stage of the progenitor star and then consecutively reach circumstellar regions composed of the wind material of the previous phases such as the red supergiant and the main-sequence phase. The main-sequence wind has roughly solar abundance~\citep{2011A&A...530A.115B} while winds of evolved phases are denser and often enriched in heavy elements. Wolf-Rayet winds exhibit both dense, rapidly-expelled stellar winds and a high chemical enrichment in C, N or O. Eventually, the forward shock will reach the ambient medium that is of solar composition for isolated Galactic supernovae~\citep{2009ARA&A..47..481A} and may be enriched in super-bubbles spawned by many (massive) stellar objects~\citep[][and references therein]{1977ApJ...218..377W,2019MNRAS.490.1961E}. The forward shock would be rather slow at that time, but may still accelerate a large number of particles up to the energy band in which the composition of cosmic rays is measured. 

The reverse shock propagates through supernova ejecta and can accelerate particles, albeit only for a  limited period of time \citep{2012APh....35..300T,2013A&A...552A.102T} and with poorly defined efficiency on account of the weak magnetic field \citep{2005A&A...429..569E}. The various evolutionary phases of the progenitor star imply substantial variation in the properties of the winds \citep{langer_araa_50_2012}, and so one must expect a complex structure of the wind zone, in particular if the progenitor moved \citep{meyer_2014bb}, is very massive~\citep{2020MNRAS.493.3548M} 
and/or if the interstellar medium is highly magnetized~\citep{2017MNRAS.464.3229M}, leading to asymmetric supernova remnants  \citep{borkowski_apj_400_1992,2015MNRAS.450.3080M,2015A&A...584A..49V}. Whenever the supernova shock hits a discontinuity in the circumstellar medium, it splits into a transmitted and a reflected shock, and so many shocks will roam the ejecta, the wind zone, and the interstellar medium beyond \citep{2007ApJ...667..226D}. Quite a few of them will be weak because they propagate in hot gas, but they may still contribute to GeV-band cosmic rays.

The enrichment of material in OB associations and the wind zone of the progenitors of core-collapse
supernovae affects the composition of both the cosmic rays accelerated
in the system and the target material for inelastic nucleus-nucleus collisions. 
Hadronic interactions in such enriched media would produce secondary particles, whose spectra 
can significantly differ, both in shape and normalization, from the spectra produced 
in the pp collisions \citep[e.g.][]{2005ApJ...620..244K,2006PhRvD..74c4018K,2009PhRvC..79c7901N,2014PhRvD..90l3014K} that dominate secondary production in the interstellar medium, because the
multiplicity spectra of secondary particles are different \citep{2007APh....27..429H,2008APh....29..282H}. Simple enhancement factors apply only far from any spectral structure like the pion bump or spectral cut-offs \citep{1970Ap&SS...6..377S,2009APh....31..341M,2016APh....81...21M}. Secondary particles like pions, but also $(\Sigma^\pm,~\Sigma^0)$, $(K^\pm,~K^0)$, and $\eta$ particles decay into radiation products such as gamma rays and neutrinos. Their spectra thus potentially allow the measurement of the composition of cosmic rays inside their sources, as well as that of the ambient material. Therefore, 
it is important to have a precise quantitative description of the spectral characteristics of their production as function of the elemental composition of the energetic particles and the cold target gas. 

In this work, we use Monte-Carlo event generators, namely {\sc DPMJET}-III \citep{roesler2001monte} and UrQMD \citep{1998PrPNP..41..255B, bleicher1999relativistic} to calculate 
inelastic cross sections and differential production rates of final-state secondary particles produced in nuclei collisions, using hydrogen (H), helium (He), carbon (C), and oxygen (O) both as projectiles and as target material. The results of these simulations are used then to calculate the gamma-ray emission from supernova remnants (SNRs) using the particle acceleration code RATPaC \citep{2012APh....35..300T, 2013A&A...552A.102T, 2016A&A...593A..20B}. We consider SNRs evolving in different environments created by stellar winds of progenitor stars featuring different composition and discuss the impact of heavy nuclei on the resulting gamma-ray spectra.


\section{Calculation of production of secondary particles}
\label{sec:secondary_production}
In particle physics, the Monte Carlo simulation approach is widely applied to obtain information on particle production in hadronic interactions. The principle advantage of the so-called event generators is that distributions for all secondary products are obtained simultaneously and self-consistently, conserving quantities like energy, baryon number, and strangeness. For astrophysical applications it is expedient to precompute the yield of secondary particles and their decay products for a given type and energy of the colliding hadrons. 

Following \citet{2007APh....27..429H,2008APh....29..282H}, we use the recently updated event generator {\sc DPMJET}-III-19.1\footnote{\url{https://github.com/afedynitch/DPMJet}} \citep{2001ICRC....2..439R,Fedynitch:2015kcn} to calculate the inelastic cross section, $\sigma_j$, and the multiplicity matrices, $\mathbb{M}_{i,j}$, for gamma rays, neutrinos, and secondary electrons and positrons. We separately consider H, He, C, and O nuclei as projectiles and as targets, leading to 16 combinations that can be arbitrarily combined to represent a wide range of abundances for cosmic rays and the
ambient medium. We follow the decay chain of unstable secondaries down to the final, stable products. Since the cosmic-ray propagation time is higher than the life time of neutrons, we also follow the decay of neutrons. As stable particles we consider ${\rm p},~{\rm e}^{\pm},~\gamma,~\nu_{\rm e},~\bar{\nu}_{\rm e},~\nu_{\mu},~\bar{\nu}_{\mu}$, for which we produce multiplicity matrices. The dominant source of gamma rays is the decay of neutral pions, while most high-energy neutrinos and ${\rm e}^{\pm}$ are produced in the decay channel $\pi^{\pm}$ to $\mu^{\pm}$ and $\mu^{\pm}$ to ${\rm e}^{\pm}$.

The spectral production rate of particle species $f$ produced in collisions of projectile particles of type $s$ with cold target nuclei of type $T$ can be defined as
\begin{equation}    
    Q_{f,s,T} \left( E \right) = \frac{\mathrm{d}n_f}{\mathrm{d}t \,  \mathrm{d}E \, \mathrm{d}V} = n_{T} \int \mathrm{d}E_{\mathrm{CR}}\ N_{\mathrm{CR},s}\left( E_{\mathrm{CR}} \right) c \beta_{\mathrm{CR}}  \sigma_{s,T}\left( E_{\mathrm{CR}} \right) \left(\frac{\mathrm{d}n_{f}}{\mathrm{d}E} \right)_{s,T} \, .  
    \label{eq:1}
\end{equation}
Here $N_{\mathrm{CR},s}$ is the differential density of the cosmic-ray species in question, \mpo{$E_\mathrm{CR}$ the total energy per cosmic-ray nucleon, $E$} is the energy of the secondary, $n_{\mathrm{T}}$ is the number density of the target nuclei of interest, $\sigma$ is the inelastic cross section, and $dn_{f}/dE$ is the multiplicity spectrum of the secondary particle $f$. 

For a binned cosmic-ray spectrum and a binned spectral
production rate, the integral in equation~\ref{eq:1} simplifies to a matrix operation that projects a vector, the cosmic-ray spectrum, onto another vector, the spectral production rate of the secondaries. The production integral can be rewritten as

\begin{equation}
    \begin{split}
            Q_\mathrm{f,s,T} \left( E_{i} \right) & = n_{{T}} \sum\limits_{j} \Delta E_{{j}} N_{\mathrm{CR},s}\left( E_j \right) c \beta_{j} \sigma_{s,T}\left( E_{{j}} \right) \left(  \frac{\mathrm{d}n_{f}}{\mathrm{d}E_i} \left( E_{i},E_j \right) \right)_{s,T} \\
            & = n_{{T}} \sum\limits_{j} \Delta E_{{j}} N_{\mathrm{CR},s}\left( E_j \right) c \beta_{j}  \sigma_{s,T}\left( E_{{j}} \right) \mathbb{M}_{i,j,{f,s,T}}\ ,
    \end{split} 
\end{equation}
where $\mathbb{M}_{i,j,{f,s,T}}$ is a matrix, or array, that carries the full information of secondary production through all relevant channels. It can be very efficiently applied to arbitrary cosmic-ray spectra at the expense of a pre-defined binning of particle energy.

\mpo{We use a log-spaced grid for the total cosmic-ray energy per nucleon, $E_\mathrm{CR}$, subdividing a range from the pion-production threshold (1.24 GeV) up to 100 PeV into 374 bins with central energy
\begin{equation}
E_j = 1.24 * 1.05^{j} \frac{\mathrm{GeV}}{\mathrm{nucleon}} , \quad j = 0, \dots, 373 \ .
\end{equation}  
The grid for secondary particles covers the range from 10 MeV up to 100 PeV subdivided into 200 bins with central energy
\begin{equation}
E_{i} = 0.01 \cdot 1.121376^{\left( i+0.5 \right)} \ \mathrm{GeV}, \quad  i = 0, \dots , 200 \ .
\end{equation}
In sources with very high UV/X-ray photon density a cosmic ray in the PeV band or at higher energy can produce gamma rays, neutrinos, and secondary electrons through p-$\gamma$ interactions \citep[e.g.][]{2017ApJ...843..109G}. These processes are not considered here.  }
 

At energies below a few GeV/nucleon, nuclear effects such as Fermi motion and binding energies become relevant, rendering several high-energy approximations in {\sc DPMJET} invalid. The Glauber model \citep{shmakov1988diagen}, which is at the basis of heavy-ion simulations with {\sc DPMJET}, has to be replaced with a model that tracks the motion of the individual nucleons.

\anatoli{In collisions involving nuclei}, secondary pions and heavier mesons are created off \anatoli{a single or multiple bound nucleons. Because multiple nucleons can fragment and produce pions, particle yields in proton-nucleus collisions are asymmetric with significantly more particles produced in the direction of the nucleus when viewed in the center of mass frame. At the same time, nuclear medium effects can strongly modify particle yields \citep{Barr:2006fs}.  If mesons hadronize within the nuclear volume, they can interact with the surrounding nucleons, initiate intranuclear cascades, leading to tertiary emission and to nuclear excitation.} In this case, mesons may be absorbed inside the nucleus by processes like $\pi^- + \text{p} + \text{p} \to \text{n} + \text{p}$ resulting in a significant suppression of particle yields compared to the proton-proton case. Also, the pion production threshold can move to lower energies and smear out due to the Fermi motion of nucleons with a characteristic energy $\langle E_\textrm{F} \rangle\sim50$ MeV. For a fixed total energy of the nucleus, each interacting nucleon pair will have slightly different initial-state kinematics leading to a less sharp pion production threshold that moves to lower energies per nucleon since the motion vectors of colliding nucleons can be oriented toward each other. \anatoli{On amplitude level, the nuclear medium affects the sharpness of the pion production threshold due to the widening of nucleonic resonances. The} fact that half of the projectile nucleons are neutrons changes the $\mathrm{e}^+/\mathrm{e}^-$ and the $\nu/\bar{\nu}$ ratio since due to isospin symmetry the production of $\pi^+$ off protons is equal to that of $\pi^-$ off neutrons.

We employ the {\sc UrQMD} 3.4 code \citep{1998PrPNP..41..255B} at low energies by linearly interpolating the multiplicity spectra of secondaries between {\sc DPMJET} and {\sc UrQMD} between 6 and 13 GeV/nucleon projectile energy. {\sc UrQMD} performs a more sophisticated simulation of the kinetic motion of nucleons and should therefore be more reliable for the reproduction of the effects mentioned above.

The decay of secondary particles into stable final particles, for example gamma rays, neutrinos, and antiparticles, can be analytically followed with standard methods \citep{1963JGR....68.4399J} or handled internally in each of the Monte-Carlo codes. The latter substantially
improves the accuracy of the spectra, as the method of Jones relies on double integrals for each decay, which can introduce substantial errors for binned spectra. We verified that internal handling gives essentially the same result as the \emph{a posteriori} calculation for very fine binning and let {\sc DPMJET}-III and {\sc UrQMD} compute all decays. We also allow for a sufficiently high number of collisions to keep the statistical uncertainty of simulations below 1\%. 

\mpo{Both {\sc DPMJET}-III and {\sc UrQMD} have been extensively tested with accelerator data, where available. Of particular interest for this paper are the p-C collision studies at moderate projectile energy conducted with the HARP experiment. The measured double-differential pion-production cross section for proton-carbon collisions is well reproduced by the event generators, at least in the region of the astrophysically relevant bulk of the distribution \citep{2008APh....29..257H}. The model modifications of the current DPMJET-III-19.1 compared to the version used for the HARP paper are minor, at least for the present application, and the comparison is still valid.}

\begin{figure}[t]
\includegraphics[width=0.9\textwidth]{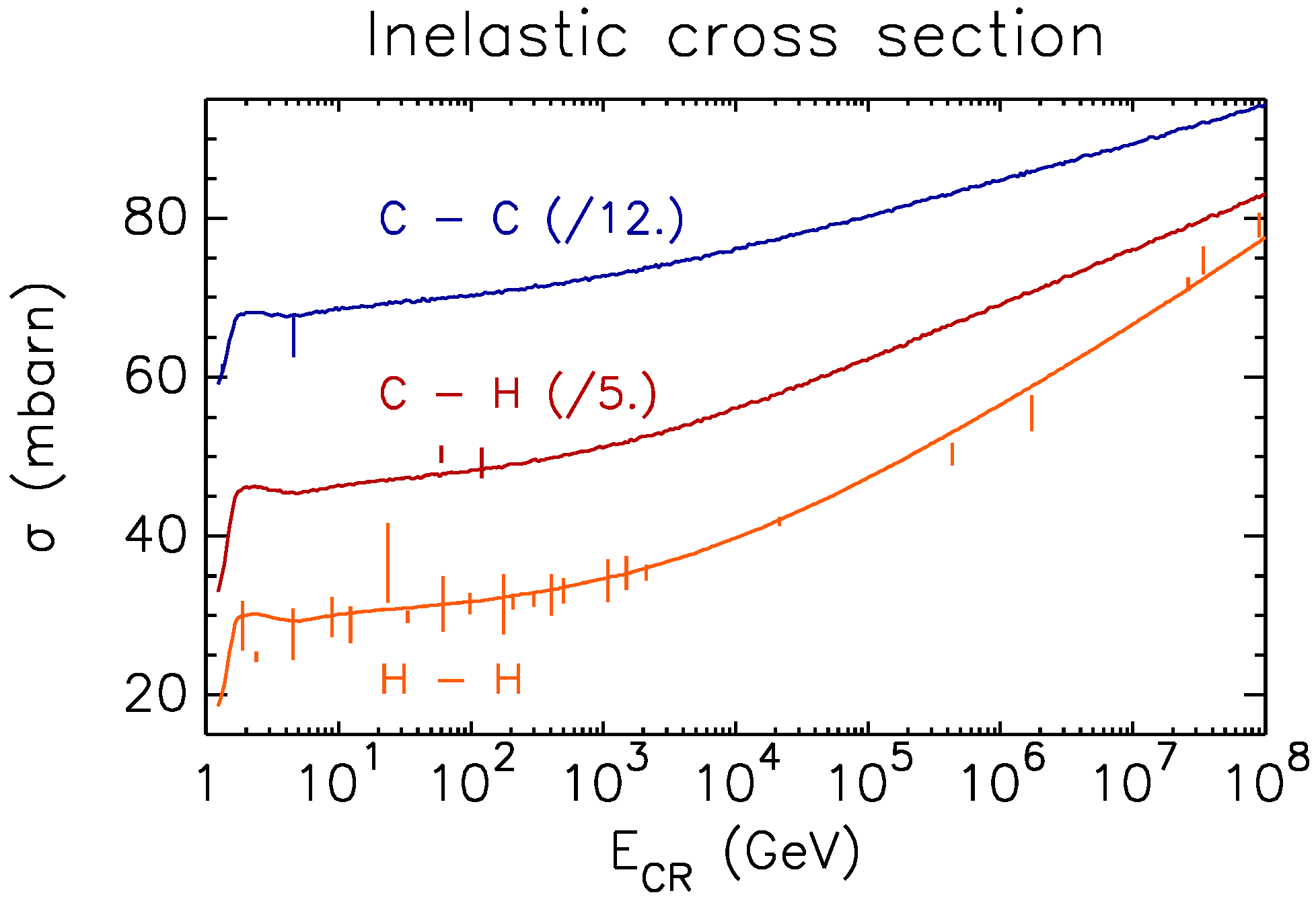}
\caption{Inelastic cross section as function of the total energy per nucleon of the projectile. C-H stands for carbon projectiles on a hydrogen target, and likewise for C-C and H-H. The cross section of the C-H and H-C collisions is the same. Scaling is applied as noted in the figure to ease visual inspection of the energy dependence. Also shown as error bars are measurement data taken from the literature.
}
\label{crosssection}
\end{figure} 

\subsection{The inelastic cross section}

For the same mass density, heavy elements have a
much lower number density than, e.g., protons. The
inelastic cross section is typically much larger though, and often the cross section is simply scaled as power of the mass number, $A$ ($\propto A^{0.7}$ or similar). Figure~\ref{crosssection} displays the inelastic cross section for combinations of hydrogen and carbon nuclei as function of the total energy per nucleon of the projectile, $E_\mathrm{CR}$. \mpo{Also shown in the figure are literature values for the inelastic cross section \citep{Aksinenko:1980nm,Antchev:2013haa,Terashima:2014wca,
Aaboud:2016mmw,Aaboud:2016ijx,Tanabashi:2018oca,Antchev:2017dia} that are generally in good agreement with the simulated cross section.} Other species combinations show a qualitatively similar behavior. 

We find that there is no simple scaling with mass number. At 10 GeV/nucleon  the cross-section ratio for C-H and H-H collisions is $A^{0.82}$, but that between C-C and C-H is only $A^{0.51}$. We also observe a significantly weaker energy dependence of the cross section for heavy collision partners than for light ones since nuclear cross sections in the Glauber model are strongly related to the nuclear geometry and less on the nucleonic interaction probability.

\section{General spectral characteristics of secondary particles}

\subsection{Gamma-rays}
\label{gamma-rays}
The characteristic turnover of the gamma-ray spectrum near a GeV is often used
to identify a hadronic origin of the emission from SNRs
such as W44 and IC443 \citep{paper:w44_ic443_pi0_fermi}. Likewise, the cut-off in the gamma-ray spectrum can be used to infer whether or not a certain SNR is a PeVatron. The shape of the underlying spectrum of cosmic rays would depend, however, on the composition of accelerated particles as well as the composition of the target material. Therefore, we cannot make definite conclusions without knowing the composition.

We shall first describe the typical impact of heavy nuclei on the resulting gamma-ray spectrum. For that purpose, we shall use a generic cosmic-ray spectrum that follows a power law in momentum with an exponential cut-off,
\begin{equation}
    N_\mathrm{CR}(p_\mathrm{CR}) = N_0\, p_\mathrm{CR}^{-s} \exp{\left(-\frac{p_\mathrm{CR}}{Z\,p_\mathrm{c}}\right)} , 
    \label{eq:6}
\end{equation}
and set the spectral index to $s=2.2$. The choice of a cut-off momentum reflects the expectations that particle transport and acceleration scales with rigidity and hence with charge number, $Z$. Here $p_\mathrm{c}$ denotes the cut-off momentum for hydrogen nuclei. Equation~\ref{eq:6} can be then rewritten in total energy per nucleon, $ E_\mathrm{CR}$, as
\begin{equation}
N_\mathrm{CR} (E_\mathrm{CR})= N_0\, A^{-1.2}\, c^{1.2}\, E_\mathrm{CR} \,\left(E_\mathrm{CR}^2 -m_p^2c^4\right)^{-1.6}\,
\exp\left(-\frac{A}{Z}\frac{E_\mathrm{CR}}{E_c}\right),
\label{CRspec_E}
\end{equation}
where we assumed the cut-off is at much higher energies than the rest mass and again $A$ is the atomic mass number.

\begin{figure}
\includegraphics[width=0.9\textwidth]{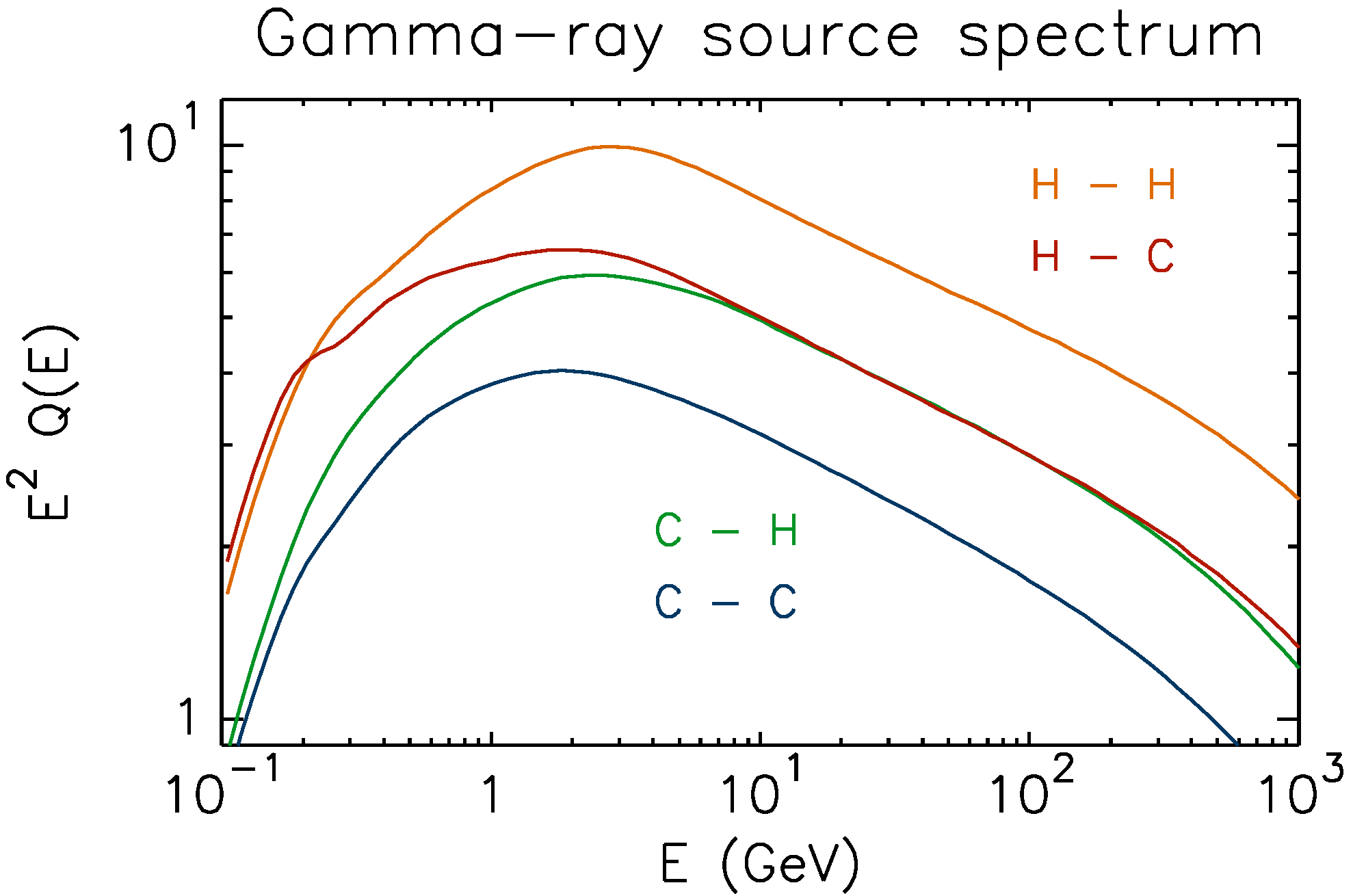}
\caption{GeV-band gamma-ray spectra for various compositions of cosmic rays and ambient gas. The cosmic-ray spectra follow Eq.~\ref{CRspec_E}. }
\label{gammaray_spec_GeV}
\end{figure}

\begin{figure}
\includegraphics[width=0.9\textwidth]{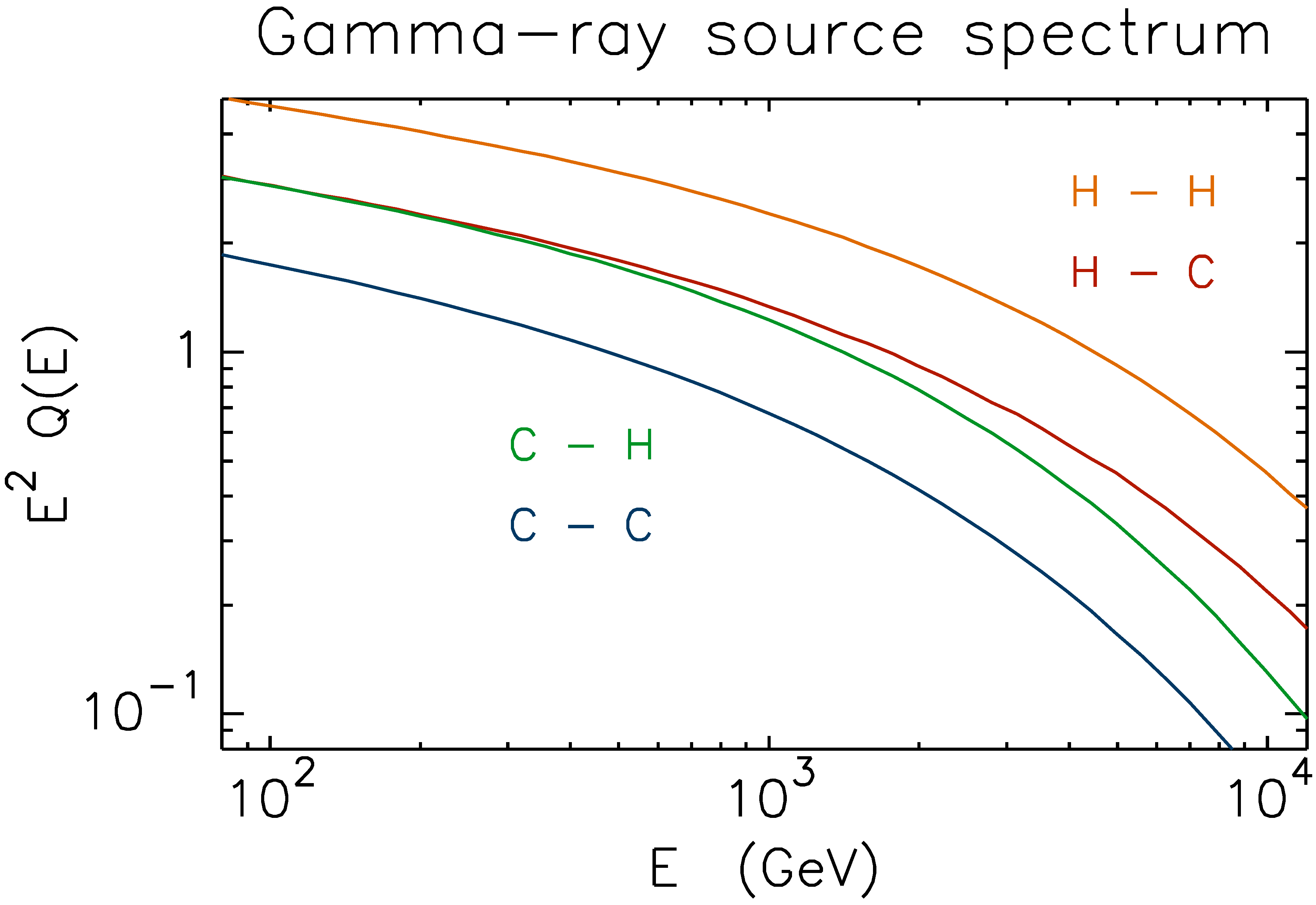}
\caption{TeV-band gamma-ray spectra for various compositions of cosmic rays and ambient gas. The cosmic-ray spectra follow Eq.~\ref{CRspec_E} with $E_c=50\,\mathrm{TeV}$. }
\label{gammaray_spec_PeV}
\end{figure}

In Figures \ref{gammaray_spec_GeV} and \ref{gammaray_spec_PeV} we separately show the resulting GeV-band and TeV/PeV-band gamma-ray spectra calculated for the fiducial cosmic-ray spectrum (Eq.~\ref{CRspec_E} with $E_c=50\,\mathrm{TeV}$) for different compositions of cosmic rays and ambient gas. We construct four scenarios with cosmic rays and target material consisting of either exclusively Hydrogen or exclusively Carbon. The figure shows the correct normalization of spectra for a fixed mass density of target material, hence $n_T \propto A^{-1}$, and a fixed number of cosmic rays in spectra starting at a fixed \mpo{momentum. If the particle spectrum started at a fixed total kinetic energy, there would be an additional scaling factor $N_0\propto A^{(s-1)/2}=A^{0.6}$} \citep{1993A&A...270...91P}. As the number density of gas atoms falls off with mass number, this assumption corresponds to an injection efficiency that is linear in $A$. The purpose of that assumption is to bring the curves closely together for better visualization of the spectral differences.

In the GeV band we note from Figure \ref{gammaray_spec_GeV} a substantial difference in the gamma-ray spectra below $5$~GeV for different types of cosmic rays and target material\anatoli{. As discussed in Section \ref{sec:secondary_production}, the angular distribution of $\pi^0$ produced in H-C collisions is asymmetric, and many pions are emitted co-moving with the fragmentation region of carbon nuclei. A detailed discussion of the differences between H-H and H-C collisions can be found, for example, in \citet{Barr:2006fs}. Therefore, more low-energy pions that result from, e.g., N$^\star$ resonances are produced almost at rest when the carbon nuclei serve as target. The bump at $200$~MeV in the H-C curve requires pions with $E_\pi\gtrsim 220$~MeV which is commensurate with that expected from the decay of the lowest nucleonic resonances. This enhancement is absent in the C-H case, where the fragmentation region of carbon is boosted, moving most pions to higher energies. Furthermore, the result is affected by nuclear medium effects such the widening of resonances, final state interactions leading to intra-nuclear cascades and the Fermi motion of the nucleons inside the nuclei that can shift the pion production threshold to lower energies. The same effect is observed in the gamma-ray spectra of H-O collisions.} Generally, heavy material will shift the peak in the spectrum toward lower energies. We find a peak energy of $2.9\,\mathrm{GeV}$ for H-H interactions, as opposed to $2.4\,\mathrm{GeV}$ for C-H, $2.0\,\mathrm{GeV}$ for H-C, and $1.8\,\mathrm{GeV}$ for C-C. 

The most prominent feature is high flux of $E\lesssim 500$~MeV gamma rays for light projectiles and heavy targets (H-C). This component would stand out, as it is much stronger than the non-thermal bremsstrahlung emission that might appear with soft spectrum in this band and would allow a determination of the magnetic-field strength in the remnant \citep{1980MNRAS.191..855C}. 

Although the systematic uncertainties in the spectral fitting of \emph{Fermi}-LAT data below a few hundred MeV can be substantial, the composition-related differences in the hadronic gamma-ray spectra in the GeV band should offer a new avenue for the tomography of particle acceleration in bright supernova remnants.

Recently, \citet{2019arXiv190908484B} showed that the volume-integrated particle spectrum in old SNRs might feature a break at 10-100~GeV due to a rapid decrease of the maximum energy reachable with shock acceleration at later times. Below the break energy the spectrum follows the usual power law with index $s = 2.0$ resulting from diffusive shock acceleration, and above that energy the power-law index is $s\approx 2.7$. This spectral shape roughly corresponds to that observed from old SNRs such as IC 443 and W44 \citep{paper:w44_ic443_pi0_fermi}. The location and shape of the turnover in the gamma-ray spectrum depends on the break energy and spectral index of the cosmic-ray spectrum and on the composition of the cosmic rays. For heavy nuclei the maximum in the gamma-ray spectrum shifts to lower energies, i.e. the break energy in the cosmic-ray spectrum corresponds to a lower energy in gamma-rays comparing to the gamma-ray spectrum obtained in pp collisions. 

In Fig.~\ref{gammaray_spec_PeV} we display gamma-ray spectra around the cut-off. The cosmic-ray spectrum again follows Eq.~\ref{CRspec_E} with cut-off energy $E_C=50$~TeV for protons and rigidity scaling for heavier nuclei. For that choice we would observe the gamma-ray cut-off in the TeV band where imaging atmospheric Cherenkov telescopes offer very high sensitivity over a wide spectral band. Apart from the normalization there is little difference in the gamma-ray spectra for light and heavy target material. However, we clearly observe a cut-off at lower gamma-ray energies for a heavy composition of the cosmic rays that reflects the cut-off scaling in cosmic-ray energy per nucleon, not the total energy of particles.  
\subsection{Neutrinos}
\mpo{Neutrinos are produced in the same hadronic interactions as gamma rays and often hailed as smoking gun of hadron acceleration. Recently, the association of a 120-TeV neutrino with a gamma-ray bright AGN led to considerable excitement \citep{2018Sci...361.1378I} and a flurry of modelling activity \citep[e.g.][]{2019NatAs...3...88G,2020ApJ...891..115P}. The very low inelastic cross section of p-$\gamma$ interactions and the high threshold energy, for pion production by a proton and a 7-eV photon about 10 PeV, render p-$\gamma$ interactions potentially dominant for PeV-scale neutrino production in AGN, GRBs, or other sources of similarly high UV/X-ray photon density \citep{2008PhRvD..78c4013K}. In a variety of other sources, like supernova remnants, generally in the Galaxy \citep{2017ApJ...849...67A}, or for TeV-scale neutrinos, nucleus-nucleus interactions are likely the dominant source process for high energy neutrinos. Here we only describe the neutrino yield in nucleus-nucleus interactions.}

\begin{figure}
\includegraphics[width=0.9\textwidth]{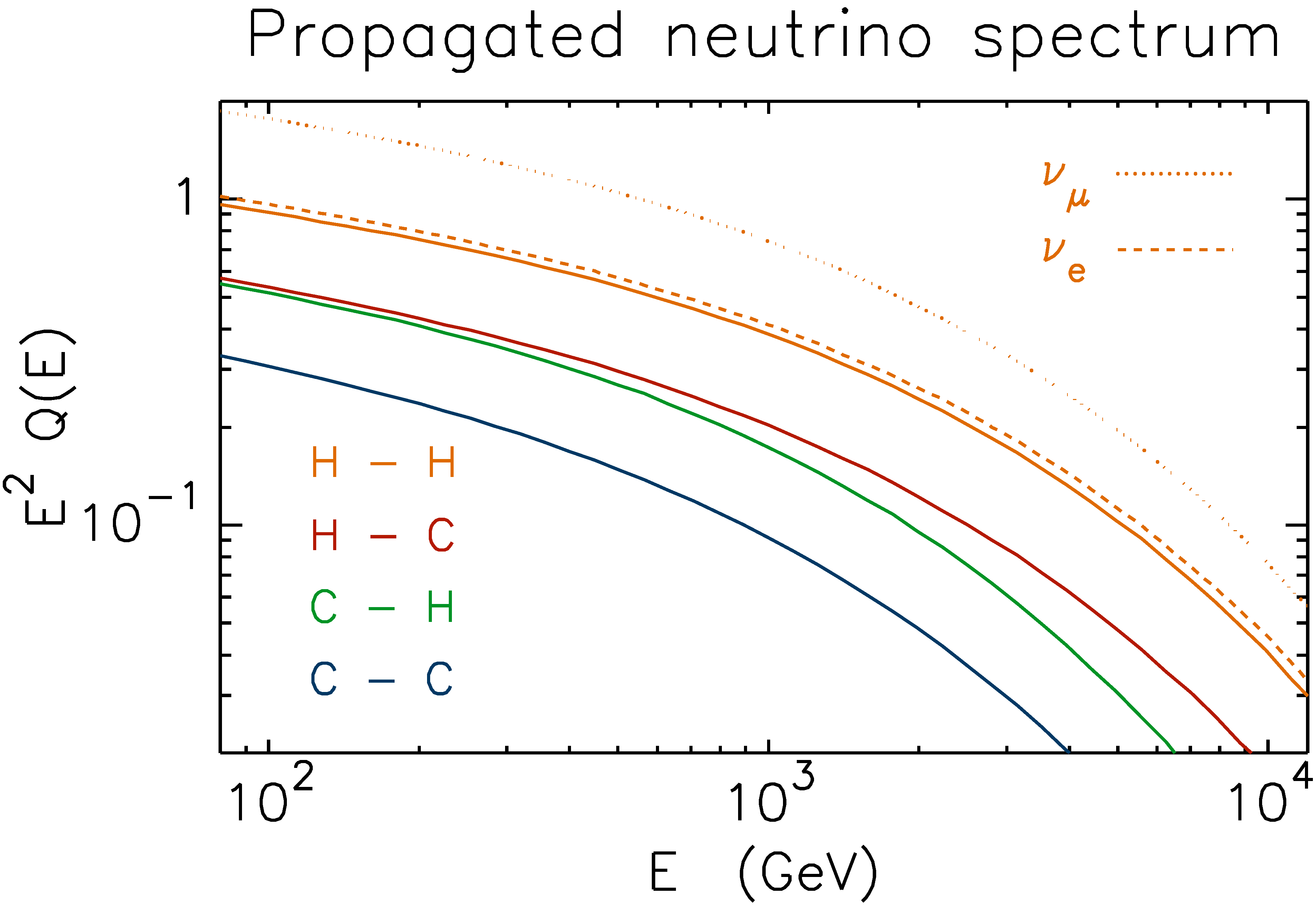}
\caption{\mpo{TeV-band neutrino spectra for various compositions of cosmic rays and ambient gas. The cosmic-ray spectra follow Eq.~\ref{CRspec_E} with $E_c=50\,\mathrm{TeV}$. Solid lines represent fully mixed neutrino spectra, i.e. $(Q(\nu_\mu)+Q(\nu_e))/3$. For proton-proton collisions we also plot as dotted and dashed lines the original production spectra for $\nu_\mu$ and $\nu_e$.}}
\label{mu_spec_PeV}
\end{figure}

\mpo{In high-energy hadronic interactions, the initial flavor ratio of neutrinos from pion decay is $\nu_e:\nu_\mu:\nu_\tau=1:2:0$. The neutrino oscillation wavelength, $\lambda_{\nu\nu} = (4\pi E_\nu/\Delta m^2) \hbar c\lesssim 10^{-3}\ $pc for PeV neutrinos \citep{PhysRevD.86.013012}, is short enough that the finite size of neutrino sources and the finite energy resolution of detectors cause an observed ratio of 1:1:1, i.e. full mixing. }

\mpo{In Figure~\ref{mu_spec_PeV} we show neutrino spectra in the cut-off region for the same setup as was used to calculate the gamma-ray spectra displayed in Figure~\ref{gammaray_spec_PeV}. The spectra are propagated, i.e. fully mixed. We also present the original source rates of $\nu_\mu$ and $\nu_e$ for light composition. To be noted from the figure is the small difference between the spectra for each flavor. Hence propagation essentially only reduces the $\nu_\mu$ flux by a factor $2$.  Also to be noted is that the neutrino spectrum in the cut-off region shows the same behaviour as that in the gamma-ray spectrum. The cut-off energy is a bit lower than that of the gamma rays, and the scaling with composition is virtually identical.}


\subsection{Positrons}

\begin{figure}[h]
\includegraphics[width=0.9\textwidth]{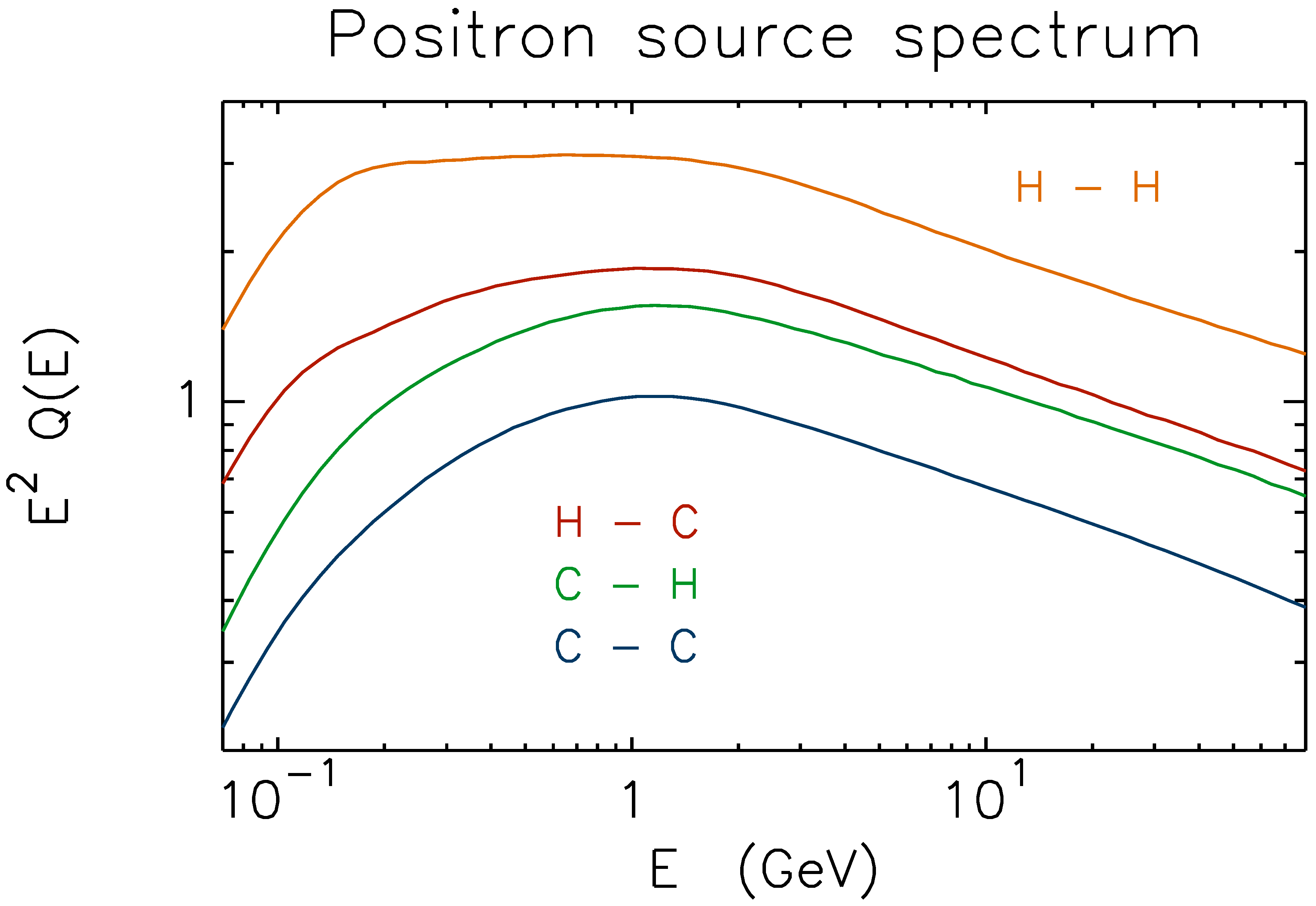}
\caption{Positron production spectra for different composition of the cosmic rays and target material. The spectrum of underlying cosmic-rays is assumed to follow Equation~\ref{CRspec_E}.}
\label{positron_spec}
\end{figure}

The multiplicity spectra and charge ratio of charged pions are substantially changed for heavy elements on account for nuclear-medium effects and the presence of neutrons in the projectile and target.
In Figure~\ref{positron_spec} we display positron spectra that are calculated for the token cosmic-ray spectrum shown in equation~\ref{CRspec_E} for different compositions of the cosmic rays and target material, as was done for gamma rays in the previous section. It is evident that for heavy projectiles we observe relatively few positrons below 1 GeV, and the typical kinematic turnover at 200 MeV exists only for pp collisions.

Secondary electrons and positrons are believed to make a substantial or even dominant contribution to the synchrotron radio emission from starburst galaxies \citep[see, e.g.,][]{1994A&A...287..453P,2009ApJ...698.1054D}. For the magnetic-field strength typically estimated for the starburst cores, $B\gtrsim 100\,\mathrm{\mu G}$, radio emission below $1$~GHz is produced by electrons or positrons at energies below 1~GeV. Using pp reaction rates, \citet{1994A&A...287..453P} demonstrated that the radio spectrum observed from M82 could be as well explained assuming secondary electrons only as with conventional primary electrons. The corresponding gamma-ray emission from hadronic interactions that one would expect in the former scenario was later observed at the required level \citep{2010ApJ...709L.152A}. In both cases a substantial opacity for free-free absorption was required. Naturally there are substantial uncertainties in propagation models for starburst cores, arising from the unknown diffusion properties and escape by advection in the wind, that can lead to some variation in the total particle spectrum of secondary electrons in M82. It remains obvious though that a heavy elemental composition in the starburst core would at least lessen the need for a high free-free opacity.

\section{Applications to SNR}

Massive stars undergo various evolutionary phases prior to their final explosion as a supernova, therefore, the expanding shock wave will eventually interact with their pre-shaped circumstellar medium. This particularly affects the elemental abundances of the remnant, which evolve time-dependently as the shock wave expands into the progenitor's surroundings \citep{2012A&A...540A.144S,2015A&A...579A..75T}. 
Core-collapse SNRs are one of the possible channel of the circumstellar evolution of high-mass stars, usually ending their lives during either a red supergiant (RSG) or Wolf-Rayet (WR) phase of evolution. In this scenario, the remnant expands into the circumstellar medium of the supernova progenitor, i.e. a blown-up wind bubble in the case of a static object, whose composition is fixed by the (i) past surface properties of the progenitor and (ii) the internal structure of the defunct star at the moment of the explosion. In this work, we consider four different stellar surroundings: that of a RSG star and three other environments corresponding to several types of WR stars, WN (Nitrogen rich), WO (Oxygen rich), and WC (Carbon rich). The corresponding elemental abundances are listed in the Table \ref{tab1} \citep{2012A&A...540A.144S,2015A&A...579A..75T,2017A&A...605A..83D,2019A&A...621A..92S}. We calculate the expected  gamma-ray emission for these four cases and compare it to the scenario in which a supernova blastwave expands into the wind zone of a main-sequence massive star, i.e. a medium of typical ISM composition. For the sake of comparison, we keep the wind mass density constant when changing its composition, hence, the dynamics of the blastwave is similar for all five cases while their emission properties differ from each other. 

\begin{table}
\begin{center}
\begin{tabular}{ l| c c c c c }
\hline
 \hline
 Components& ISM&  RSG& WN& WO& WC\\
 \hline
 Hydrogen & 0.71 & 0.639 & 0.20 & 0.0 & 0.0\\
 Helium & 0.28  & 0.349  & 0.78 & 0.14 & 0.55\\
 Oxygen & 2.06e-3 & 5.41e-3 & 0.0 & 0.24 & 0.05\\
 Nitrogen & 0.0 & 3.1e-3 & 1.5e-2& 0.0& 0.0 \\
 Carbon & 2e-3 & 1.42e-3 & 1e-4 & 0.62 & 0.4\\
 Iron & 4e-4 & 1.35e-3 & 1.4e-3 & 0.0 & 1.6e-3\\
 \hline
\end{tabular}
\caption{Mass fractions of certain elements assumed for the wind models considered here.}\label{tab1}
\end{center}
\end{table}

\subsection{Simulation setup}

To calculate the gamma-ray emission from the SNR we use the code RATPaC (Radiation Acceleration Transport Parallel Code) which is described in detail elsewhere \citep[and references therein]{2018A&A...618A.155S,2019arXiv190908484B,2019A&A...627A.166B}. The code simultaneously solves the transport equation for cosmic rays, the transport equation for magnetic turbulence,
and the hydrodynamic equations for the gas flow. The equations are solved time-dependently in one dimension under the assumption of spherical symmetry. Hydrodynamic simulations of the shock evolution are performed using the {\sc pluto } code \citep{2012ApJS..198....7M} which is on-the-fly incorporated into RATPaC.
In this study, we are interested in describing the impact of the composition of the ambient material on the resulting gamma-ray spectrum from the SNR, for which we can use a simplified setup. Instead of solving the transport equation for magnetic turbulence we assume Bohm diffusion in the vicinity of the shock. 

To isolate the effect of elemental composition, we use the same hydrodynamic setup for all five cases ignoring the differences in the wind parameters for the different types of star. For the stellar wind we assume a mass-loss rate of $\dot{M_\mathrm{w}} = 3\times10^{-5}M_\odot$/yr and a wind velocity of $V_\mathrm{w} = 2500$\,km/s which are typical for a WR wind~\citep{2000A&A...360..227N,2015A&A...578A..66T}. We also assume that the density in the wind zone decreases as $1/r^2$ with distance from the star, i.e. the wind speed is constant as a function the distance to the star, and that the termination shock of the circumstellar medium is distant enough -- or equivalently the SNR sufficiently young -- to ensure that the supernova shock wave is still expanding into the unperturbed wind. We run our simulations for 1000 years, after which the radius of the forward shock is $10.7$~pc. The mass of the swept-up stellar wind material is about $0.13~M_\odot$.

We parametrize the circumstellar magnetic field with a $1/r$ profile,
\begin{equation}
    B(r) = B_\ast  \frac{R_\ast}{r},
\end{equation}
where $B_\ast$ ($=10\ $G) is the magnetic-field strength at the surface of the star and $R_\ast$ ($=100\,R_\odot$) is the radius of the star, respectively. The magnetic field is assumed to be compressed at the shock by a factor of $\sqrt{11}$ and passively transported downstream. To calculate the magnetic-field profile downstream we solve the induction equation for ideal MHD.

For particle injection into diffusive shock acceleration we adopt the thermal leakage model \citep{2005MNRAS.361..907B}, a parametrization that relates the number of particles to the population in the high-energy tail of the downstream Maxwellian. Above a certain injection momentum, $p_\mathrm{inj}$, all particles are supposed to be able to cross the shock back to the upstream region for further acceleration. The injection momentum is given as multiple of the thermal momentum, $p_{\mathrm{inj}} = \xi p_{\mathrm{th}}$, and the injection parameter is set to $\xi = 4.2$. Particles are injected at every time step at the shock position with momentum $p_{\mathrm{inj}}$.


\subsection{Results}

To better than 10\% accuracy the high-energy
cosmic-ray spectra of different elements scale with the
total energy divided by the charge number, because that reflects the
rigidity for relativistic particles. The scaling in
the normalization of spectra is more complex than the
simple mass-number correction of the mass density in
the gas fluid to the number density scaling relevant for
injection into shock acceleration, because most
injected particles reside at low energies near the
injection energy. For the same injection energy, e.g.
the same multiple of the downstream temperature, $kT$,
the injection rigidity scales with the mass number and
the charge number as $R\propto \sqrt{A}/Z$.

\begin{figure}[h]
\includegraphics[width=\textwidth]{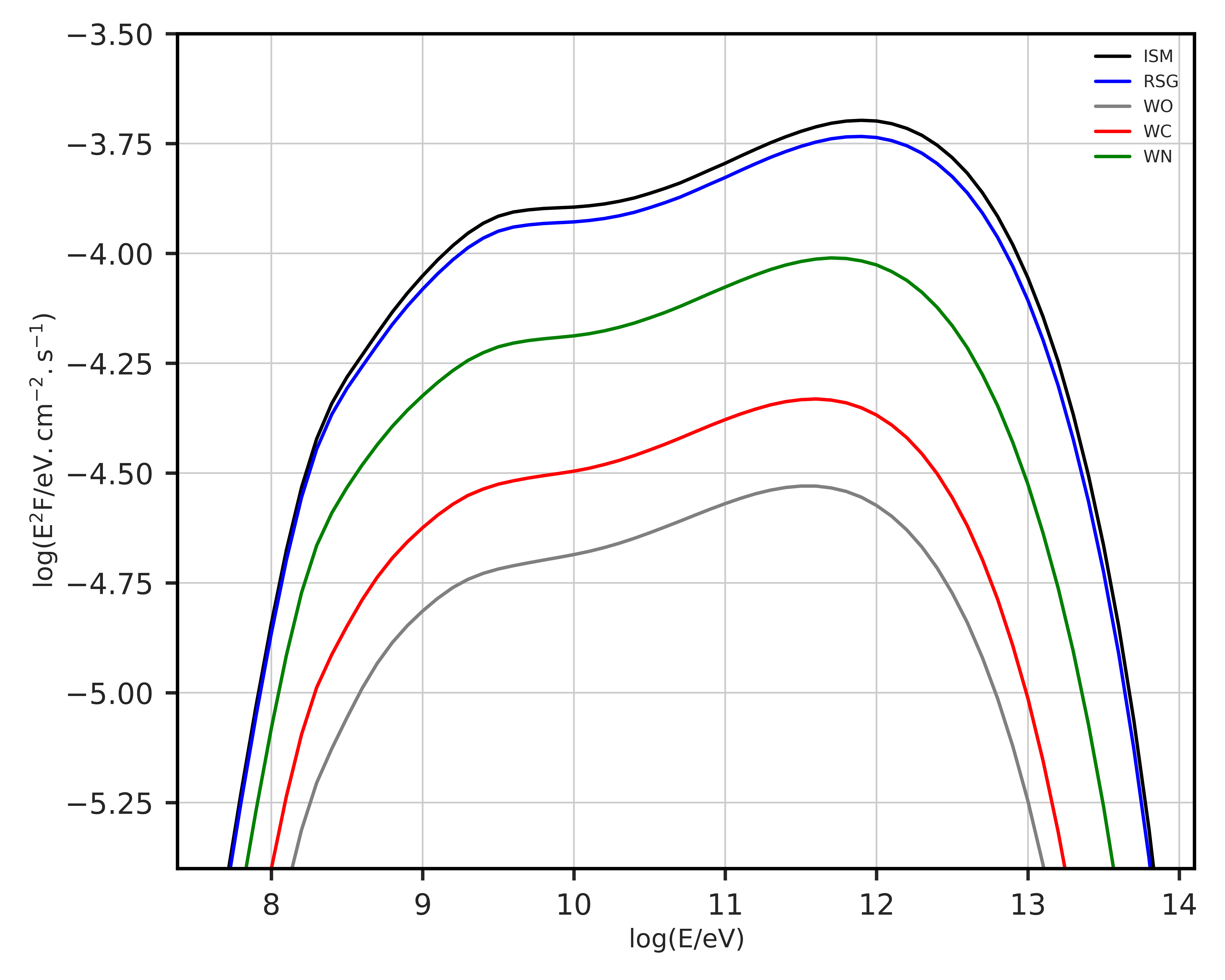}
\caption{Simulated gamma-ray spectra from a core-collapse SNR at an age of 1000 years for five stellar models with different wind composition. The elemental abundances for these models are listed in Table~\ref{tab1}.\label{snr_gammaray_spec}}
\end{figure}

In Figure~\ref{snr_gammaray_spec} we show the expected gamma-ray spectra from an SNR that is 1000 years old and expanding in five environments with different composition as detailed in Table~\ref{tab1}. The underlying cosmic-ray spectra reflect the entire acceleration history of the remnant and the properties of the ambient medium. Particles enter the acceleration from the thermal pool passing through the shock, and hence cosmic rays have the same composition as the target material. 

The gamma-ray spectra are not normalized, and their amplitude reflects the impact of the composition on the difference in normalization of the resulting gamma-ray emission. The large difference in normalization compared to Figures~\ref{gammaray_spec_GeV} and \ref{gammaray_spec_PeV} arises because plasma of fixed mass density has a number density that inversely scales with the mass number of the dominant element. The number density of cosmic rays scales with the number density of the plasma from which they are accelerated.
As a consequence, SNR evolving in a wind zone with heavy composition (WO, WC, and WN) would generate fewer gamma-rays (Fig.~\ref{snr_gammaray_spec}). SNR evolving in a RSG wind 
produce a similar gamma-ray spectrum as SNR expanding in a medium with ISM composition. The slight difference in the H and He abundances does not significantly impact the spectra.

As already shown in Section~\ref{gamma-rays}, the cut-off energy of the gamma-ray spectrum shifts to lower energies for heavier composition which can also be seen for the stellar-wind models we considered here (Fig.~\ref{snr_gammaray_spec}). This implies that the observed gamma-ray spectrum of an SNR would correspond to a higher maximum energy of cosmic rays, if the cosmic rays are heavy nuclei, potentially pointing to particles accelerated to PeV energies. However, the downside is that we expect considerably fewer gamma-ray photons from heavy nuclei, which naturally makes the detection harder. The forthcoming Cherenkov Telescope Array (CTA) has a significantly higher sensitivity than current instruments and can potentially open a new window of opportunities in this domain.

Similarly, at GeV energies, where the gamma-ray emission can be observed by the \emph{Fermi}-LAT, the impact of the composition on the shape of the low-energy turnover in observations might be obscured by a significantly lower gamma-ray flux for heavier compositions. For SNRs interacting with a dense medium the gamma-ray flux would still be high, and the knowledge of the composition would be essential to properly interpret the spectra.

To be noted from Fig.~\ref{snr_gammaray_spec} are the hard spectra above 30 GeV. They are not related to cosmic-ray feedback or similar, but reflect the inner structure of the remnant that can not be reproduced with simple one-zone models. TeV-scale cosmic rays have a larger mean free path than GeV-scale cosmic rays, and so they far more efficiently penetrate the interior of the remnant, where the cosmic-ray spectrum is therefore unusually hard. In the deep interior the gas density is high on account of the $r^{-2}$ density profile in the progenitor's freely-expanding stellar wind and so is the gamma-ray emissivity. In total, very-high-energy cosmic rays radiate more efficiently than GeV-band cosmic rays do. 

\section{Summary}

Core-collapse SNRs evolve in the wind-blown bubbles of their progenitors. The elemental composition of the circumstellar medium significantly depends on the type of the progenitor star. For Wolf-Rayet stellar winds the ambient gas would be much heavier than for main-sequence or RSG winds, featuring a high abundance of Carbon, Nitrogen and Oxygen. 

We calculated the inelastic cross sections and production matrices for secondary gamma rays, neutrinos, and electrons/positrons generated in hadronic interactions for heavy elements, namely He, C and O, using the Monte-Carlo event generators {\sc DPMJET-III} and {\sc UrQMD}. We find that the inelastic cross section does not simply scale with mass number and that there is a substantial change in the energy dependence of the cross section for heavy elements compared to, e.g., protons. Simulated cross sections and matrices are then used to calculate the hadronic gamma-ray emission from core-collapse SNRs evolving in wind zones with different compositions. \mpo{We do not consider leptonic radiation processes in this paper.}

Our simulations show that heavy elements among cosmic rays and in the ambient medium substantially change the resulting spectrum of gamma-rays in SNRs. For the same injection efficiency into diffusive shock acceleration and a given mass density of gas passing through a shock, cosmic rays of heavy composition are fewer, and so they produce a low gamma-ray flux. Below 
3~GeV, the spectral peak  and the shape of the turnover in the gamma-ray spectrum strongly vary with the composition of both the cosmic rays and the ambient medium. This complicates the search for the nonthermal bremsstrahlung component at energies between $100$~MeV and $300$~MeV. Nevertheless, precise measurements of the gamma-ray emission of the hadronic origin from SNRs can potentially be used to probe the elemental composition of their environments.

One the high-energy side of the gamma-ray spectrum, the maximum photon energy 
decreases with heavier composition for a rigidity scaling of the underlying spectrum of cosmic rays. This implies that for a heavier composition the observed gamma-ray spectrum from an SNR would reflect a higher maximum energy of cosmic rays. Besides, the fewer gamma rays emitted by SNRs evolving in a wind zone with heavy composition would also make them harder to detect. 

Our results for SNRs can be extrapolated to, e.g., gamma-ray production in binary systems with colliding winds or run-away star. For the former, Wolf-Rayet stars were suggested to be promising targets  \citep{2006ApJ...644.1118R}, but the estimates of the gamma-ray flux were made assuming ISM composition. Consequently, for realistic Wolf-Rayet wind compositions the luminosity estimates have to be reduced by a factor of a few. For run-away stars, the wind-termination shock was identified as dominant particle accelerator, but the shocked ISM is the most important target material for hadronic gamma-ray production \citep{2018ApJ...864...19D}. Care must be exercised to properly account for the composition of the wind material, and more so on the efficiency of diffusive transport from the shocked wind to the shocked ISM. A similar transport issue arises for old core-collapse SNR, in which particles accelerated in the Wolf-Rayet wind zone propagate to the outer shell of red-supergiant or main-sequence wind material that has solar composition.  

We also studied the effect of heavy nuclei on the production of neutrinos and positrons. In contrast to protons, heavy cosmic rays produce relatively few secondary positrons below $1$~GeV and hence do not contribute to the synchrotron radio emission below $1$~GHz. This is particularly relevant for starburst galaxies like M82, for which the bulk of the radio synchrotron emission may be produced by secondary electrons.

The cross sections and multiplicity matrices will be made available upon request. 

\section*{Acknowledgements}
AF completed his work as JSPS International Research Fellow (JSPS KAKENHI Grant Number 19F19750).



\bibliographystyle{elsarticle-harv} 
\bibliography{bibliography.bib}





\end{document}